\documentclass[a4paper,11pt]{article}
\usepackage{pos}
\usepackage{setspace}
\usepackage{subcaption}
\usepackage{doi}

\title{Advancing ARA: the Next-Generation (ARA-Next) DAQ System}
 \ShortTitle{Advancing ARA: the Next-Generation (ARA-Next) DAQ System}

\author*[a]{Pawan Giri} 
\onbehalf{for the ARA Collaboration \\{\normalsize \normalfont(a complete list of authors can be found at the end of the proceedings)}\\}
\affiliation[a]{University of Nebraska - Lincoln,\\
Lincoln, NE}
\emailAdd{pgiri4@huskers.unl.edu}

\abstract{
The Askaryan Radio Array (ARA) has been operating at the South Pole for over a decade, searching for ultra-high-energy astrophysical and cosmogenic neutrinos using the Askaryan effect. ARA has consistently served as a testbed for innovative trigger designs and advancing electronic upgrades, with ongoing data acquisition (DAQ) improvements over the past 2–3 years and a long-term plan to transition to Radio Frequency System-on-Chip (RFSoC) technology. This upgrade enables real-time data processing and sophisticated triggers, enhancing efficiency by identifying double pulses from in-ice neutrino interactions, using templates for cosmic rays, searching for real-time coincidences with the IceCube detector observations, and filtering anthropogenic noise through directional analysis. In 2024, two of the five ARA stations received DAQ upgrades, improving the existing electronics, with RFSoC-based DAQ foreseen in the coming years. In this proceedings contribution, recent ARA activities are presented, with emphasis on the planned ARA-Next trigger strategies involving RFSoC technology and the 2024–2025 season upgrades of the existing ATRI-based DAQ system to its revised version.
}

\ConferenceLogo{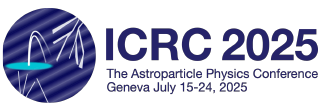}

\FullConference{39th International Cosmic Ray Conference (ICRC2025)\\
 15–24 July 2025\\
Geneva, Switzerland\\}
\begin{document}
\maketitle
\section{Introduction}
\begin{figure}[ht]
    \centering
    \begin{minipage}[t]{0.495\linewidth}
        \centering
        \includegraphics[width=\linewidth, height=5.9cm]{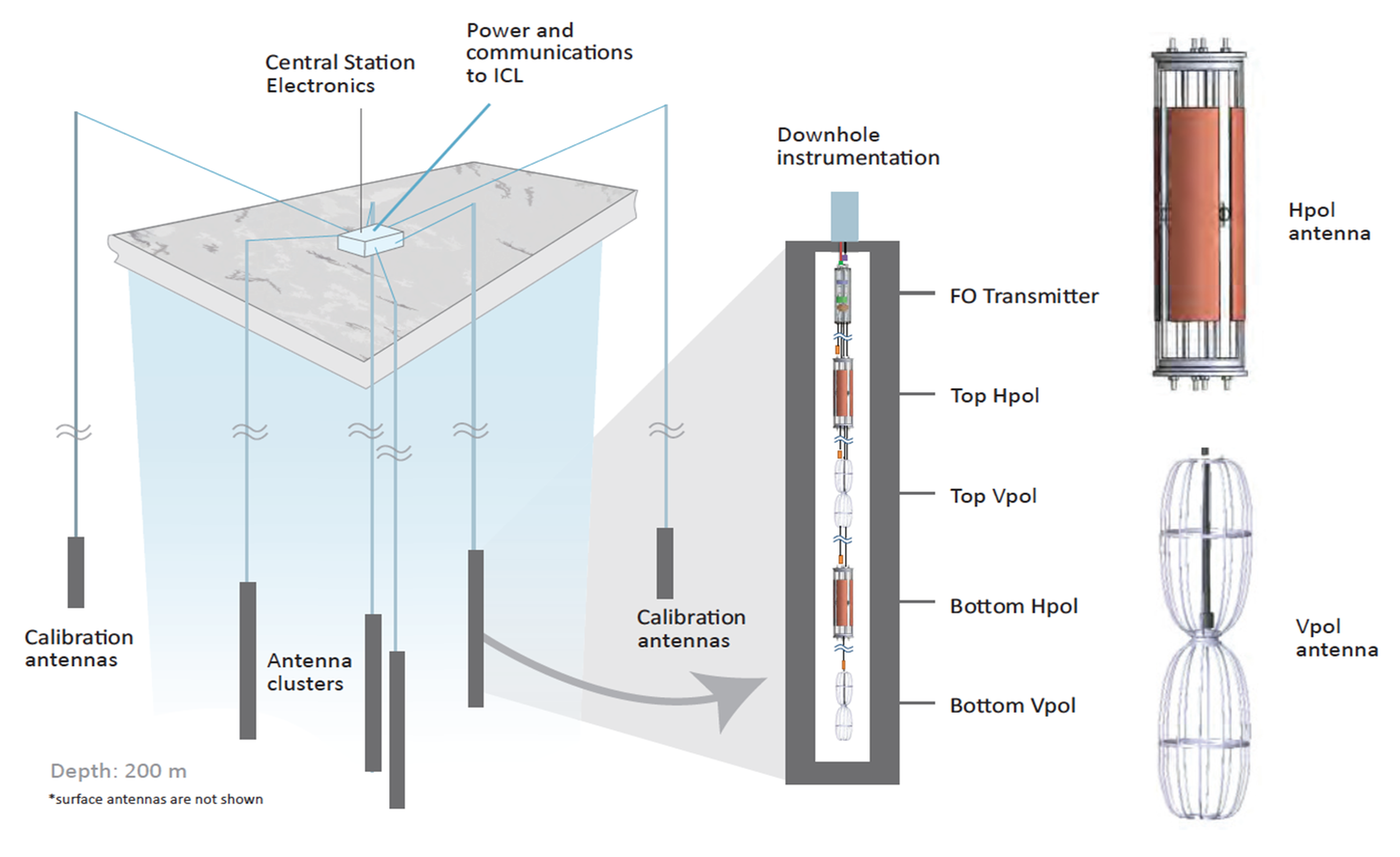}
        \subcaption{}
        \label{fig:station}
    \end{minipage}
    \hfill
    \begin{minipage}[t]{0.495\linewidth}
        \centering
        \includegraphics[width=\linewidth, height=6cm]{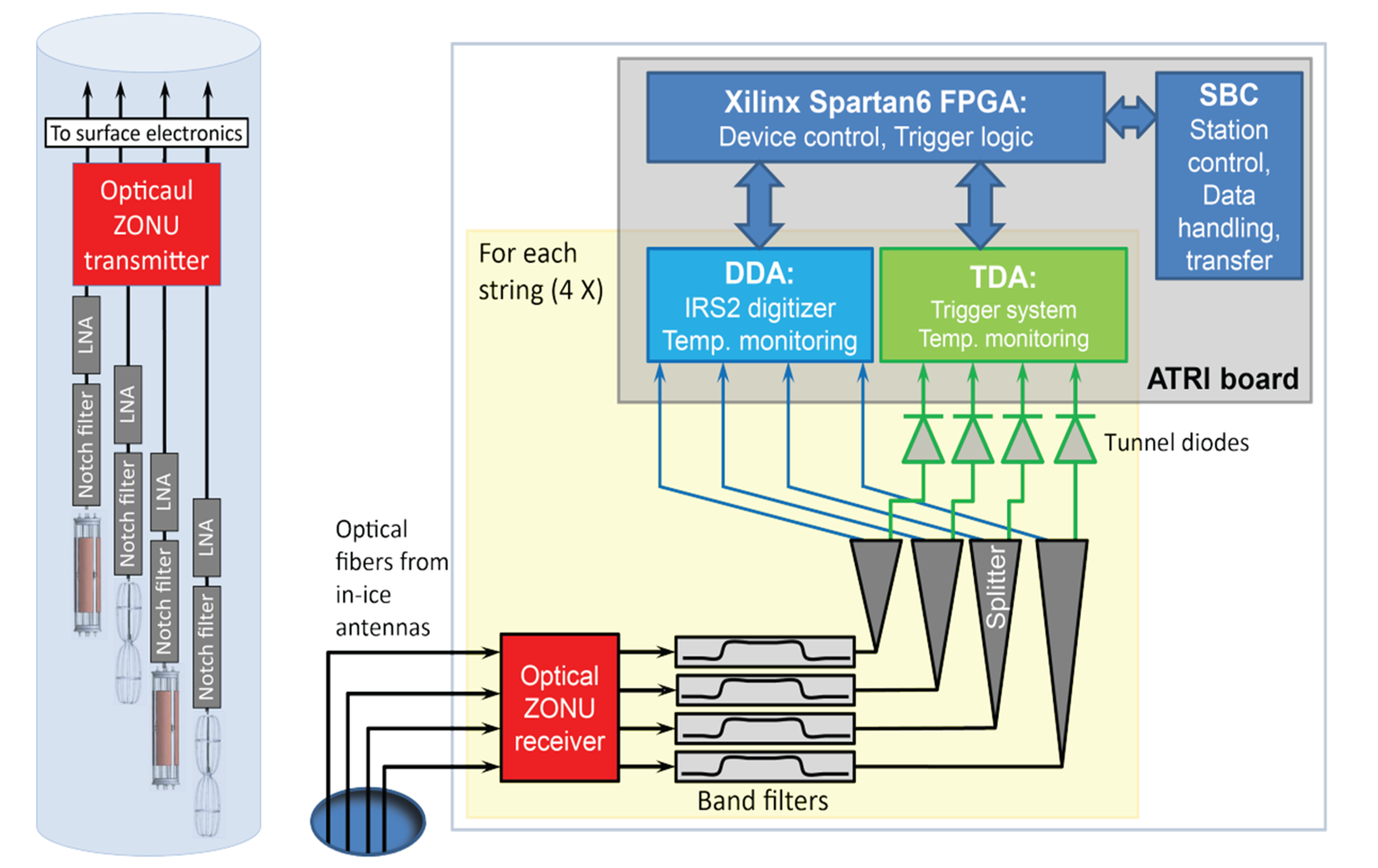}
        \subcaption{}
        \label{fig:ATRI}
    \end{minipage}
    \caption{(a) \textbf{Left}: Schematic of a traditional ARA station layout, showing four detector strings, two calibration pulser strings. \textbf{Right:} Representative designs of the VPol and HPol antennas. (b)\textbf{Left:} Components of the downhole signal chain for each string in the ARA stations. \textbf{Right:} Surface data acquisition system of an ARA station, highlighting key elements. Components outlined in yellow are shared across all strings.}
    \label{fig:Fig1}
\end{figure}
\noindent
The Askaryan Radio Array, commonly known as ARA and located at the South Pole near the IceCube Neutrino Observatory, detects neutrinos via the Askaryan effect ~\cite{Askaryan1962}, a phenomenon in which a high-energy particle cascade in a dense dielectric medium produces a coherent burst of radio-frequency Cherenkov radiation due to a net excess of negative charge. Neutrino interactions in the ice initiate such cascades, allowing ARA to probe the origin of cosmic rays, high-energy astrophysical sources, and particle interactions beyond the reach of terrestrial accelerators~\cite{Allison:2012, ARAInstr2015}. 
 A major scientific goal of ARA is to identify cosmogenic neutrino particles produced when ultra-high energy (UHE) cosmic rays interact with the cosmic microwave background through the Greisen-Zatsepin-Kuzmin (GZK) mechanism ~\cite{Beresinsky1969}. The array consists of five stations, each operating autonomously with vertically and horizontally polarized radio antennas (Vpol and Hpol) deployed in ice in a square geometry at depths of up to 200 meters (see Fig.~\ref{fig:station}).

ARA has served as a testbed for emerging technologies in neutrino astronomy. The decade-long operation of the ARA has demonstrated robust autonomous performance in extreme polar conditions, as well as long-term reliability of in-ice instrumentation and custom trigger electronics. Current efforts to modernize the data acquisition system, including the integration of RFSoC-based platforms, reflect the broader contribution of ARA to the development of scalable and efficient radio detection technologies for next-generation observatories~\cite {ARAUpgrade2023}.

\noindent
\section{Limitations of current DAQ and the RFSoC Upgrade Path}
The current ARA data acquisition (DAQ) board, known as the ARA Trigger and Readout Interface (ATRI) (see Fig.~\ref{fig:ATRI}), was developed to support the detection of radio signals in ice resulting from neutrino interactions~\cite{Allison:2012}. It uses low-noise amplifiers (LNAs) along with the IceRay Sampler 2 (IRS2) chip, a 3.2 GSa/s digitizer. Triggering is performed using fast power threshold logic with coincidence across multiple antennas. Although this architecture has enabled the successful operation of the initial ARA stations, several limitations have been identified in terms of digitization accuracy, calibration, and, more importantly, the trigger design constraints.

The forthcoming upgrade will adopt a DAQ system based on RFSoC technology, which combines RF digitizers, FPGAs, and embedded processors in a unified platform to enable high-speed integrated digitization, triggering, and data acquisition. RFSoC technology has already been adopted in various astroparticle physics and radio-frequency detection applications, including liquid scintillator experiments and phased-array radio telescopes~\cite{Axani2021,Liu2024,Pisanu2021,Zhang2024}. In particular, its low noise floor, high sampling rates, and FPGA-level logic have been instrumental in improving trigger efficiency and data throughput~\cite{Axani2023}. This proven track record reinforces the suitability of RFSoC as the core technology for the ARA-Next DAQ upgrade.

\emph{\textbf{Digitization Constraints and Sampling issues:}}
The IRS2 chip has good timing resolution, but calibration presents significant challenges. Sampling-time inconsistencies have been identified at several ARA stations, attributed to timing jitter from IRS2 fabrication variations and limitations inherent to its Switched Capacitor Array architecture~\cite{Allison:2012}. The digitizer channels for Vpol and Hpol also sample at different rates. During signal reconstruction, these sampling issues introduce complications that necessitate additional correction procedures.

The upcoming RFSoC-based DAQ board is expected to provide sub-nanosecond timing resolution and improved digitization performance. This feature of the electronic board is important for detecting the fast, broadband Askaryan radio pulses from UHE neutrino interactions in Antarctic ice. RFSoC-based digitizers have demonstrated improved spectral response and reduced RMS noise levels in similar high-precision applications~\cite{Axani2023, AMD2024}.

\emph{\textbf{System Flexibility and Reprogrammability Limitations:}}
The ATRI DAQ system includes a Xilinx Spartan FPGA, but the system remains fundamentally limited by the fixed-function IRS2 digitizers, which prevent dynamic reconfiguration of trigger logic during operation and lack the reprogrammability needed for adaptive signal processing. This inflexibility presents a challenge for future ARA upgrades, where implementing and evaluating advanced trigger strategies is a key objective.

RFSoC platforms are reprogrammable, which will allow ARA to upgrade firmware and adapt the trigger logic or signal processing pipelines without hardware changes. Flexibility is especially important for long-term experiments such as ARA, where evolving physics goals and background conditions may necessitate updates to DAQ processing, particularly given the logistical difficulty of performing physical upgrades at a remote site like the South Pole. The multichannel architecture of the RFSoC, along with its ability to support synchronized, high-throughput signal processing across multiple channels and boards, makes it well suited for future station upgrades or deployment in denser arrays~\cite{iWave2022}.

\emph{\textbf{Triggering Limitations and Real-Time Logic:}}
The current ARA trigger logic works in four stages. It starts by checking if any individual antenna sees power above a set threshold and ends by requiring that three of the eight antennas with the same polarization exceed the power threshold within a short time window~\cite{ARAUpgrade2023}. Because this triggering scheme is relatively simple, it is more susceptible to triggering on background signals such as thermal fluctuations or anthropogenic noise, which are not associated with neutrino interactions, as noted in~\cite{Allison:2012}. The current system performs real-time coincidence triggering but lacks the flexibility to implement more advanced algorithms, such as those incorporating pulse shape analysis or inter-antenna signal coherence, due to its fixed, hardware-defined logic.

The embedded programmable logic available in the RFSoC will enable real-time implementation of advanced trigger algorithms directly in hardware, reducing reliance on external processing and supporting lower-latency, high-selectivity triggering. In the context of ARA, RFSoC can enable triggering on features such as double pulses expected from in-ice neutrino interactions or the ability to correlate reconstructed directions with other detectors such as IceCube. Such functionalities are impractical or inefficient with ATRI-based systems~\cite{ARAUpgrade2023}. Real-time filtering at the hardware level, including implementations of machine learning–based algorithms, also supports directional vetos, improving the rejection of anthropogenic backgrounds.

\emph{\textbf{Data Throughput, Storage, and Event Filtering:}}
Currently, ARA stores the triggered events locally on the disk, and most of the data transfer is deferred until the end of the Austral winter season due to the remote location of the detector~\cite{shultz2016performance}. Although the current system handles event rates and data volumes adequately, it lacks the flexibility to perform event filtering or classification prior to storage. This results in the accumulation of large volumes of background-dominated data, much of which is discarded during offline analysis. The next-generation DAQ system is being designed to reduce unnecessary data collection and improve the efficiency of both storage and transfer.

\begin{figure}
\centering
\includegraphics[width=0.9\linewidth, height=9cm]{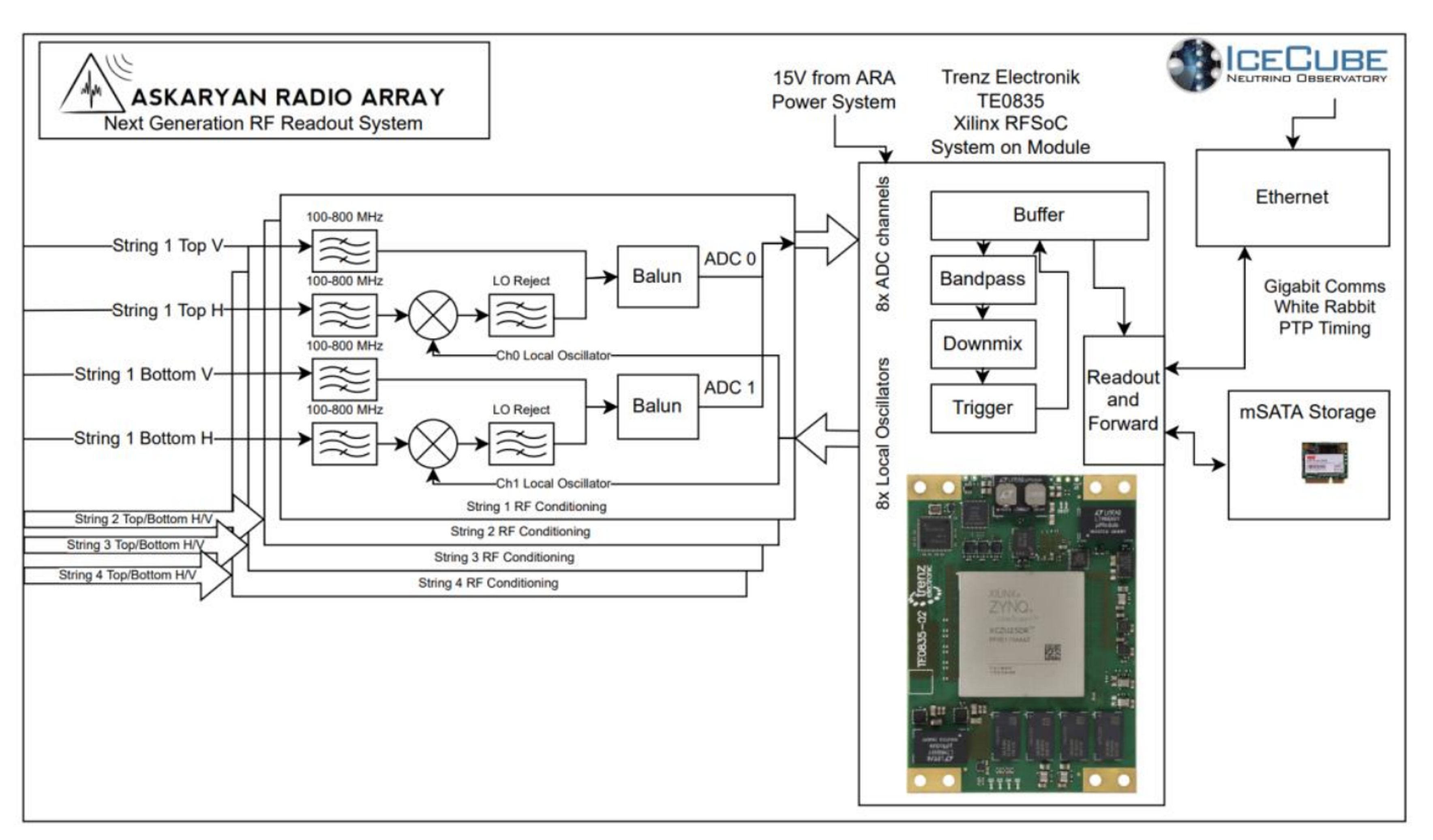}
\caption{\label{fig:Nxt_DAQ} Simplified block diagram of the ARA-Next system, illustrating 16-channel readout at an ARA station using an 8-channel RFSoC System-on-Module. Each pair of HPol and VPol inputs is combined into a single channel, with HPol signals upconverted prior to summation. The resulting 8 channels are digitized by the RFSoC, where data is buffered and processed through a trigger path. Upon satisfying trigger conditions, the buffered waveforms are read out and stored locally.}
\end{figure}

RFSoC-based DAQ designs (see Fig.~\ref{fig:Nxt_DAQ}) are expected to minimize this challenge through real-time processing capabilities that enable on-board event characterization and filtering. These features would allow ARA to keep scientifically valuable signals while reducing the storage of routine noise triggers~\cite{ARAUpgrade2023}.

\emph{\textbf{Scalability, Maintainability, and System Simplification:}}
The ARA DAQ has proven to be reliable in extreme polar conditions, but was not built with flexibility or straightforward maintenance in mind. As noted in~\cite{Allison:2012}, its tightly coupled hardware and firmware make updates harder to implement. Any necessary changes must be planned carefully within a short seasonal window, which limits how adaptable the DAQ can be over time.

The next-generation DAQ system will use fewer hardware components since RFSoC technology packs most radio functions into one chip-ADCs, DACs, digital up and down converters, ARM processors, and programmable logic all integrated together. This provides several concrete advantages. First, it is significantly more straightforward for engineers to design and maintain, substantially reducing overall system complexity. More importantly, the system enables more advanced signal processing prior to trigger decisions, while maintaining comparable power consumption to the current ATRI boards, potentially even reducing it. This is particularly important in remote Antarctic stations, where power resources are limited, and on-site maintenance is difficult or infeasible.

\section{Vision and Triggering Strategy for ARA-Next}

After years of operation, ARA has identified key aspects where the data collection process needs improvement. The Antarctic environment presents ongoing challenges, and the ARA experiment needs hardware that can keep up with evolving physics requirements. That is the driving force behind ARA-Next.

Currently, ARA digitizes everything that triggers and ships it to the North for further analysis. The goal is to handle most of the processing directly at the collection point. The future DAQ logic will look for physics in the trigger event, validating timing of the hits in different antennas, matching signals against the patterns already known, detecting cosmic ray pulses when they occur, syncing with IceCube in real time, filtering out interference from IceCube Lab, and other known sources, etc. This flexibility offers significant advantages. Furthermore, if new background event types are identified or noise characteristics evolve, trigger algorithms can be updated remotely. Real-time processing, therefore, provides substantial capability for determining data retention versus rejection, enhancing the overall efficiency of the DAQ system.

For ARA, DAQ upgrades to RFSoC-based systems open up trigger possibilities that were not previously available. Instead of just looking for signals above a certain threshold as currently implemented, the system will be able to run complex algorithms that can recognize target signatures. The goal is to detect more genuine neutrino events while reducing the contamination that currently fills the data streams. The following presents the main trigger concepts under development.

\begin{figure}[ht]
    \centering
    \begin{minipage}[t]{1.0\textwidth}
        \centering
        \begin{subfigure}[t]{0.56\textwidth}
            \centering
            \includegraphics[width=\linewidth, height=5.9cm]{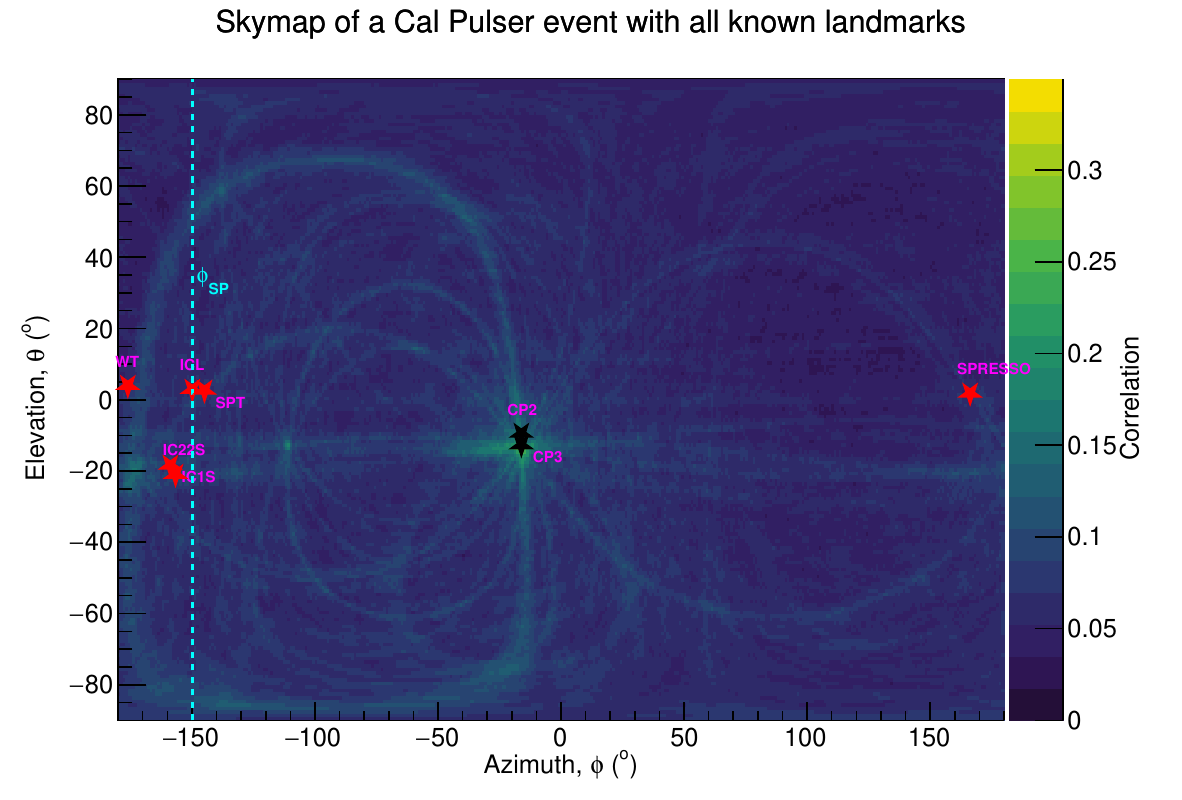}
            \caption{}
            \label{fig:skymap}
        \end{subfigure}
        \hfill
        \begin{subfigure}[t]{0.43\textwidth}
            \centering
            \includegraphics[width=\linewidth, height=5.7cm]{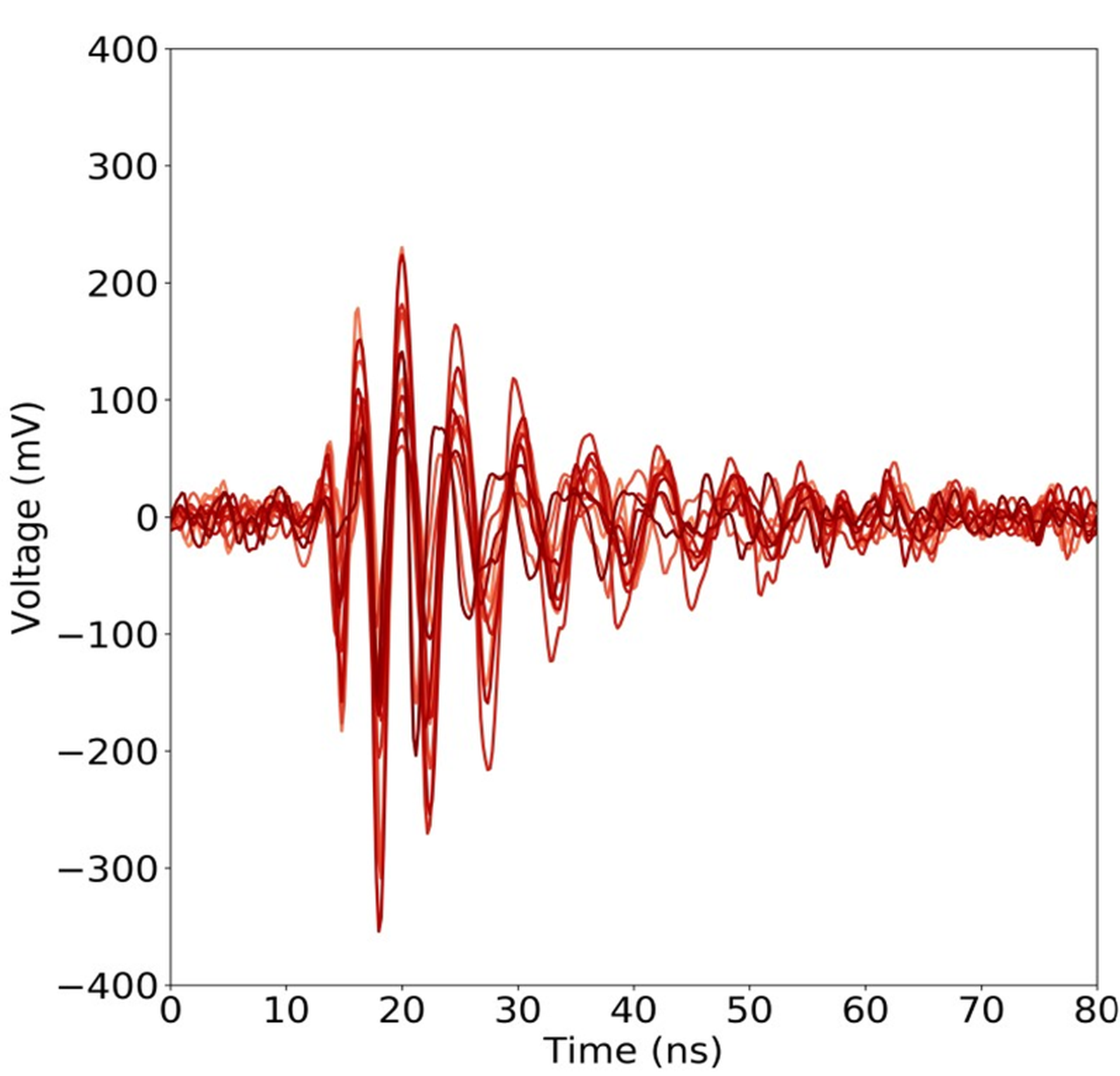}
            \caption{}
            \label{fig:CR_temp}
        \end{subfigure}
        \caption{\label{fig:trigs} (a) Skymap for ARA Station 4, showing known landmarks including the local calibration pulser (CP), IceCube Lab (ICL), South Pole Telescope (SPT), wind turbine (WT), and other locations. (b) Cosmic-ray candidates produce nearly identical waveforms that can be identified using a template-based triggering approach.}
    \end{minipage}
\end{figure}

\begin{figure}[ht]
    \centering
    \includegraphics[width=0.7\linewidth, height=6.5cm]{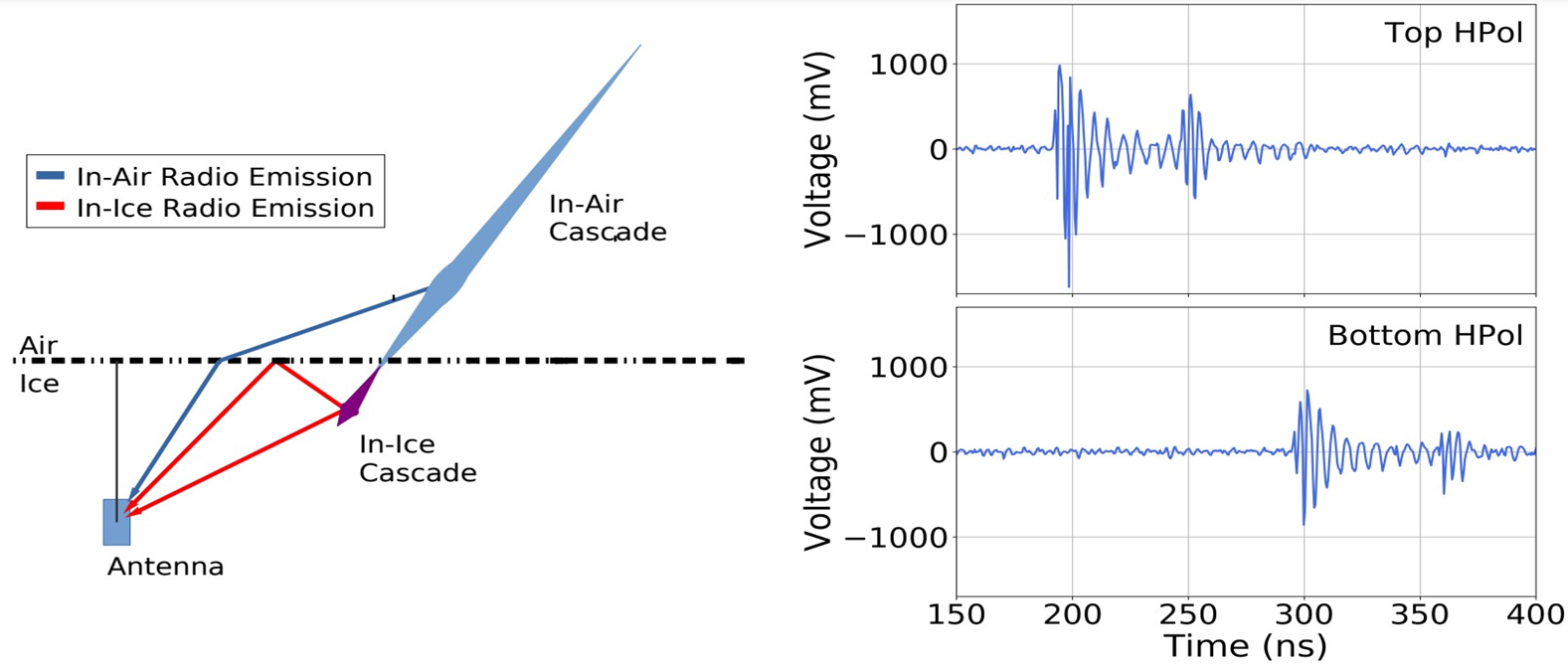}
    \caption{\label{fig:GMAS} Illustration of a candidate cosmic-ray air shower detected by ARA, where two signals separated by $\sim$50\,ns are expected from geomagnetic and in-ice Askaryan emission.}
    %\label{fig:GMAS}
\end{figure}

\emph{\textbf{Directional Triggers:}}  
Through an ongoing analysis across the entire array, ARA is mapping all the major sources of anthropogenic noise (see Fig.~\ref{fig:skymap}). The experiment can utilize beam-forming techniques, such as those already implemented with the Phased Array Station, in the regular ARA stations to determine signal arrival directions. With this directional information, triggers can be configured to reject events from surface noise or known interference sources while maintaining high sensitivity for deep, in-ice interactions expected from neutrinos.

\emph{\textbf{Template Matching Triggers:}}  
ARA-Next will use template matching to identify and reject background signals, particularly radio pulses from cosmic-ray air showers that come from above the antennas. These signals appear very similar to neutrino signatures, but have characteristic waveform patterns (see Fig.~\ref{fig:CR_temp}) that can be identified in real time using matched filtering in the firmware. The system will compare incoming signals to known templates from surface events and flag or veto triggers on the spot, cutting down background contamination without hurting sensitivity to neutrinos in the ice.

\emph{\textbf{Triggers for Geomagnetic and In-Ice Askaryan Emission:}}  
ARA has already recorded a potential candidate event for geomagnetic emission accompanied by an in-ice Askaryan pulse (see Fig.~\ref{fig:GMAS}). An innovative trigger concept can be designed to tag events originating from above the antenna center, not directed toward known landmarks, and having multiple pulses across multiple channels, with sequential hits. This trigger can simultaneously look for hits in antennas of both polarizations.

\emph{\textbf{Deep-Ice Double Pulse Triggers:}}  
One of the trigger schemes will be dedicated to deep ice neutrino events with double pulses\textemdash an initial pulse from below the antenna array (unlike cosmic-ray events, which originate above the antenna center), followed by a second pulse from above, with an arrival time gap ranging from a few nanoseconds to several hundred nanoseconds. This approach should assist in distinguishing real neutrino signals from single-pulse backgrounds such as radio interference or thermal noise; however, the trigger will be designed to avoid misidentifying single pulses from below the antenna center as background events.

\emph{\textbf{Coincidence Triggers:}}  
ARA can implement coincidence triggers at multiple levels. For interstation coincidences, the array can look for events that trigger multiple ARA stations within a short time window. This will target tau neutrino interactions, where the tau lepton from a charged-current interaction decays in flight and creates a secondary shower that downstream stations can detect. For interdetector coincidences, ARA can coordinate with IceCube to identify events that occur simultaneously in the same region of sky. A real-time coincidence trigger would use GPS timestamps to flag events that appear in both detectors within their observation windows. This will also bring ARA into the multi-messenger astronomy field.

\emph{\textbf{Machine Learning-Based Triggers:}}  
With the additional computational power available in RFSoC firmware and ARM processors, ARA plans to implement lightweight machine learning algorithms trained on simulated Askaryan pulses and actual noise data from the detectors. These ML triggers could perform real-time waveform classification, searching for characteristic features of coherent Cherenkov radiation, such as expected polarization patterns and pulse shapes from neutrino interactions in ice, that may be too subtle or complex for current threshold-based triggers to detect.

Work is ongoing for the development of a Python-based streaming trigger simulation that utilizes both real and simulated data to verify parts of the current ARA trigger and explore new design concepts. In parallel, early prototyping is underway on the RFSoC 4x2 board at the University of Nebraska–Lincoln to examine feasibility and resource requirements. Full-scale RFSoC DAQ board development for ARA-Next is in progress at Ohio State University, with the goal of deploying RFSoC-based systems at two ARA stations by the end of 2027.
\section{Pathfinding to ARA-Next: 2024--2025 Intermediate Upgrades}
\begin{figure}[ht]
    \centering
    \begin{minipage}[t]{0.63\textwidth}
        \centering
        \begin{subfigure}[t]{0.4\textwidth}
            \centering
            \includegraphics[width=\linewidth, height=3.5cm]{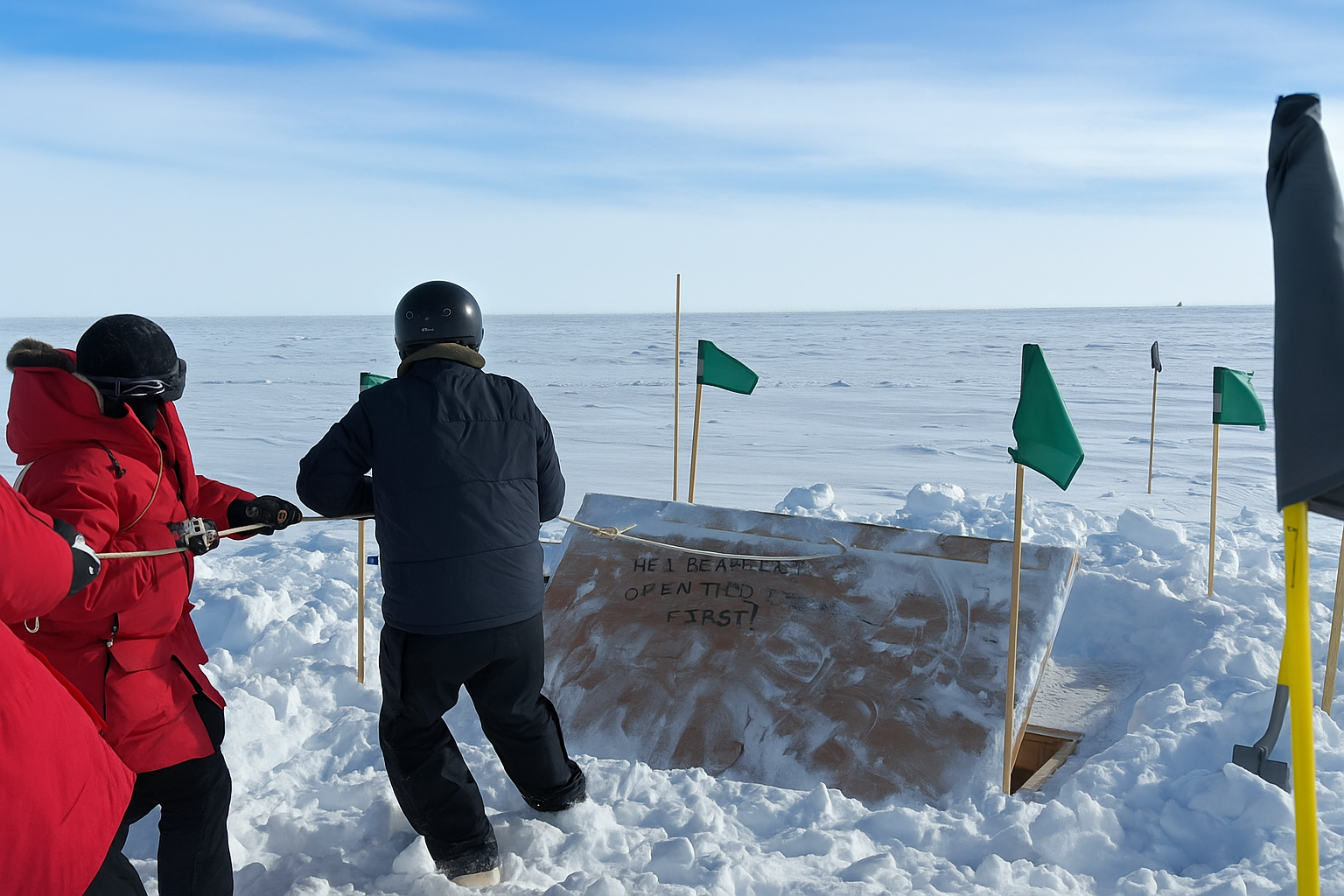}
            \caption{}
            \label{fig:fig1a}
        \end{subfigure}
        \hfill
        \begin{subfigure}[t]{0.4\textwidth}
            \centering
            \includegraphics[width=\linewidth, height=3.5cm]{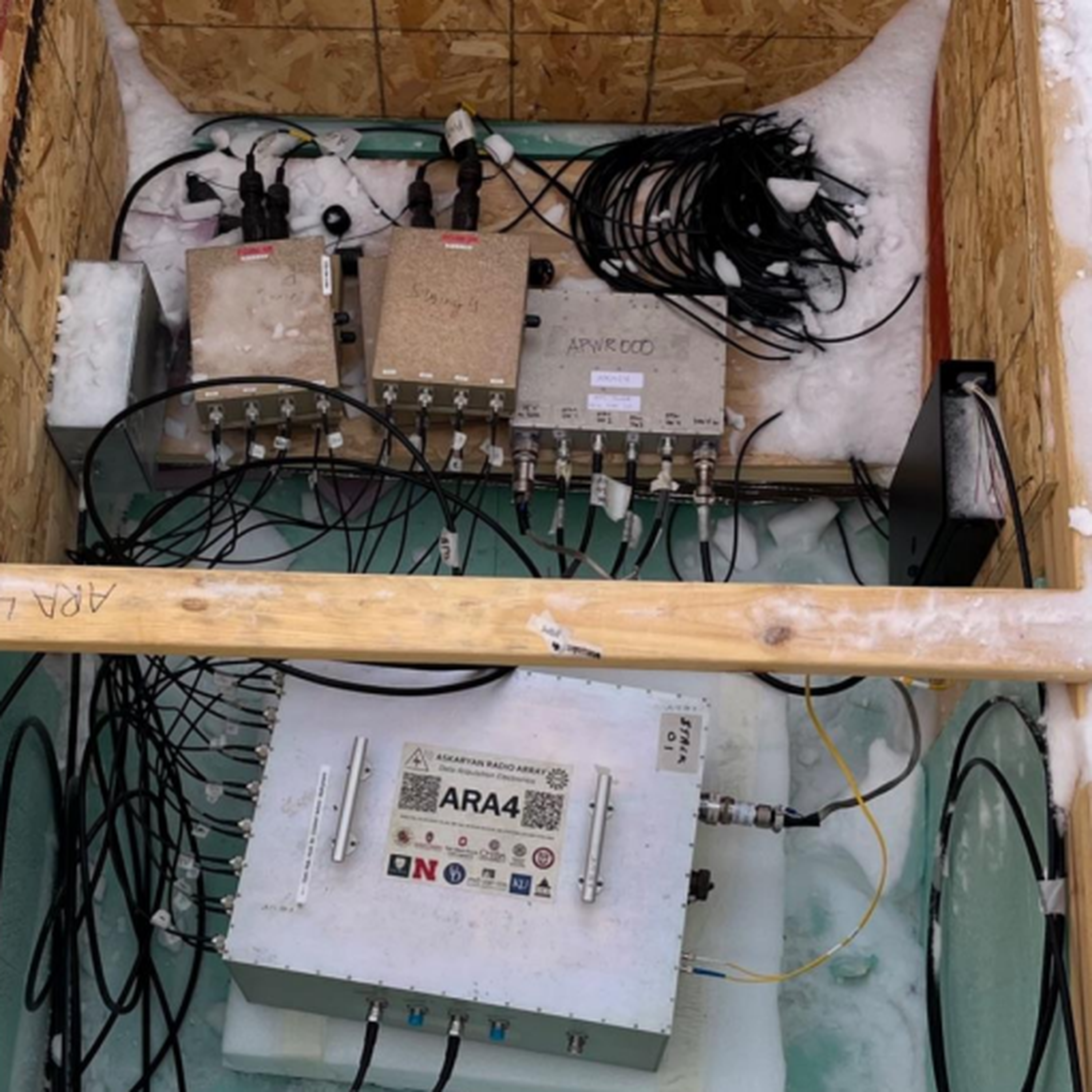}
            \caption{}
            \label{fig:fig1b}
        \end{subfigure}
    \end{minipage}
    \hfill
    \caption{\label{fig:South_pole} (a)  Maintenance team excavating the station electronics box. (b) Redeployed A4 DAQ system with upgraded ATRI system.}
\end{figure}

During the 2024–2025 austral summer, ARA reinstalled ATRI-based data acquisition boards at two of its five stations, Station 2 (A2) and Station 4 (A4), incorporating a revised USB FX2 communication interface to address persistent reliability issues observed in earlier deployments. Failures of the original FX2 controller had previously rendered both stations non-operational by disrupting critical communication pathways. Although A2 and A4 are still in a re-commissioning phase, with firmware and software undergoing updates to restore full data-taking capability, the hardware intervention successfully reestablished control and readout functionality. While independent of the RFSoC upgrade path, this effort contributed operational insights relevant to the design and implementation of the forthcoming ARA-Next data acquisition system.

\section{Summary}
ARA is preparing to transition to its next phase, ARA-Next, where all five stations will be upgraded with RFSoC-based DAQ systems. This new architecture will allow complex trigger algorithms that can detect low signal-to-noise ratio events and complex event topologies with distinct physical signatures, such as double pulses or signals originating from specific directions. The new trigger capabilities will enhance the detector’s sensitivity to targeted neutrino signals, effectively increasing instrument effective volume without deploying additional stations, which should make ARA the most effective in-ice detector in this energy range for years to come.

ARA expects to publish its ongoing analysis results in 2026, which could identify the first ultra-high energy neutrino in the radio regime or set the strongest limits on UHE neutrino flux, even before any upgrades. Once ARA implements the RFSoC-based design, we expect improvements in detection performance beyond the current standard. The sophisticated timing and processing capabilities of the ARA-Next DAQ will enable real-time coordination with the IceCube detector, based on the coincidence trigger framework described earlier. This enhanced coordination could provide the first confirmation of ultra-high-energy neutrinos through simultaneous optical and radio detection methods. The experience of the collaborative neutrino search could help design IceCube-Gen2 radio components~\cite{IceCubeGen2TDR_2024}.

%% \tableofcontents
\begingroup
\scriptsize
\setstretch{0.2}
\setlength{\bibsep}{0.02pt}
\bibliographystyle{unsrt}     % standard unsorted-by-author, numbered
\bibliography{sample}
\endgroup

%\bibliographystyle{unsrt}

%\bibliography{sample}
\clearpage
% ICRC list for ARA Collaboration
\section*{Full Author List: ARA Collaboration (June 30, 2025)}

\noindent
N.~Alden\textsuperscript{1}, 
S.~Ali\textsuperscript{2}, 
P.~Allison\textsuperscript{3}, 
S.~Archambault\textsuperscript{4}, 
J.J.~Beatty\textsuperscript{3}, 
D.Z.~Besson\textsuperscript{2}, 
A.~Bishop\textsuperscript{5}, 
P.~Chen\textsuperscript{6}, 
Y.C.~Chen\textsuperscript{6}, 
Y.-C.~Chen\textsuperscript{6}, 
S.~Chiche\textsuperscript{7}, 
B.A.~Clark\textsuperscript{8}, 
A.~Connolly\textsuperscript{3}, 
K.~Couberly\textsuperscript{2}, 
L.~Cremonesi\textsuperscript{9}, 
A.~Cummings\textsuperscript{10,11,12}, 
P.~Dasgupta\textsuperscript{3}, 
R.~Debolt\textsuperscript{3}, 
S.~de~Kockere\textsuperscript{13}, 
K.D.~de~Vries\textsuperscript{13}, 
C.~Deaconu\textsuperscript{1}, 
M.A.~DuVernois\textsuperscript{5}, 
J.~Flaherty\textsuperscript{3}, 
E.~Friedman\textsuperscript{8}, 
R.~Gaior\textsuperscript{4}, 
P.~Giri\textsuperscript{14$\dagger$}, 
J.~Hanson\textsuperscript{15}, 
N.~Harty\textsuperscript{16}, 
K.D.~Hoffman\textsuperscript{8}, 
M.-H.~Huang\textsuperscript{6,17}, 
K.~Hughes\textsuperscript{3}, 
A.~Ishihara\textsuperscript{4}, 
A.~Karle\textsuperscript{5}, 
J.L.~Kelley\textsuperscript{5}, 
K.-C.~Kim\textsuperscript{8}, 
M.-C.~Kim\textsuperscript{4}, 
I.~Kravchenko\textsuperscript{14$*$}, 
R.~Krebs\textsuperscript{10,11}, 
C.Y.~Kuo\textsuperscript{6}, 
K.~Kurusu\textsuperscript{4}, 
U.A.~Latif\textsuperscript{13}, 
C.H.~Liu\textsuperscript{14}, 
T.C.~Liu\textsuperscript{6,18}, 
W.~Luszczak\textsuperscript{3}, 
A.~Machtay\textsuperscript{3}, 
K.~Mase\textsuperscript{4}, 
M.S.~Muzio\textsuperscript{5,10,11,12}, 
J.~Nam\textsuperscript{6}, 
R.J.~Nichol\textsuperscript{9}, 
A.~Novikov\textsuperscript{16}, 
A.~Nozdrina\textsuperscript{3}, 
E.~Oberla\textsuperscript{1}, 
C.W.~Pai\textsuperscript{6}, 
Y.~Pan\textsuperscript{16}, 
C.~Pfendner\textsuperscript{19}, 
N.~Punsuebsay\textsuperscript{16}, 
J.~Roth\textsuperscript{16}, 
A.~Salcedo-Gomez\textsuperscript{3}, 
D.~Seckel\textsuperscript{16}, 
M.F.H.~Seikh\textsuperscript{2}, 
Y.-S.~Shiao\textsuperscript{6,20}, 
S.C.~Su\textsuperscript{6}, 
S.~Toscano\textsuperscript{7}, 
J.~Torres\textsuperscript{3}, 
J.~Touart\textsuperscript{8}, 
N.~van~Eijndhoven\textsuperscript{13}, 
A.~Vieregg\textsuperscript{1}, 
M.~Vilarino~Fostier\textsuperscript{7}, 
M.-Z.~Wang\textsuperscript{6}, 
S.-H.~Wang\textsuperscript{6}, 
P.~Windischhofer\textsuperscript{1}, 
S.A.~Wissel\textsuperscript{10,11,12}, 
C.~Xie\textsuperscript{9}, 
S.~Yoshida\textsuperscript{4}, 
R.~Young\textsuperscript{2}
\\\\
$^\dagger$Corresponding Authors\\
$^*$Presenter
\\\\
\textsuperscript{1} Dept. of Physics, Enrico Fermi Institute, Kavli Institute for Cosmological Physics, University of Chicago, Chicago, IL 60637\\
\textsuperscript{2} Dept. of Physics and Astronomy, University of Kansas, Lawrence, KS 66045\\
\textsuperscript{3} Dept. of Physics, Center for Cosmology and AstroParticle Physics, The Ohio State University, Columbus, OH 43210\\
\textsuperscript{4} Dept. of Physics, Chiba University, Chiba, Japan\\
\textsuperscript{5} Dept. of Physics, University of Wisconsin-Madison, Madison,  WI 53706\\
\textsuperscript{6} Dept. of Physics, Grad. Inst. of Astrophys., Leung Center for Cosmology and Particle Astrophysics, National Taiwan University, Taipei, Taiwan\\
\textsuperscript{7} Universite Libre de Bruxelles, Science Faculty CP230, B-1050 Brussels, Belgium\\
\textsuperscript{8} Dept. of Physics, University of Maryland, College Park, MD 20742\\
\textsuperscript{9} Dept. of Physics and Astronomy, University College London, London, United Kingdom\\
\textsuperscript{10} Center for Multi-Messenger Astrophysics, Institute for Gravitation and the Cosmos, Pennsylvania State University, University Park, PA 16802\\
\textsuperscript{11} Dept. of Physics, Pennsylvania State University, University Park, PA 16802\\
\textsuperscript{12} Dept. of Astronomy and Astrophysics, Pennsylvania State University, University Park, PA 16802\\
\textsuperscript{13} Vrije Universiteit Brussel, Brussels, Belgium\\
\textsuperscript{14} Dept. of Physics and Astronomy, University of Nebraska, Lincoln, Nebraska 68588\\
\textsuperscript{15} Dept. Physics and Astronomy, Whittier College, Whittier, CA 90602\\
\textsuperscript{16} Dept. of Physics, University of Delaware, Newark, DE 19716\\
\textsuperscript{17} Dept. of Energy Engineering, National United University, Miaoli, Taiwan\\
\textsuperscript{18} Dept. of Applied Physics, National Pingtung University, Pingtung City, Pingtung County 900393, Taiwan\\
\textsuperscript{19} Dept. of Physics and Astronomy, Denison University, Granville, Ohio 43023\\
\textsuperscript{20} National Nano Device Laboratories, Hsinchu 300, Taiwan\\

\section*{Acknowledgements}

\noindent
The ARA Collaboration is grateful to support from the National Science Foundation through Award 2013134.
The ARA Collaboration
designed, constructed, and now operates the ARA detectors. We would like to thank IceCube, and specifically the winterovers for the support in operating the
detector. Data processing and calibration, Monte Carlo
simulations of the detector and of theoretical models
and data analyses were performed by a large number
of collaboration members, who also discussed and approved the scientific results presented here. We are
thankful to Antarctic Support Contractor staff, a Leidos unit 
for field support and enabling our work on the harshest continent. We thank the National Science Foundation (NSF) Office of Polar Programs and
Physics Division for funding support. We further thank
the Taiwan National Science Councils Vanguard Program NSC 92-2628-M-002-09 and the Belgian F.R.S.-
FNRS Grant 4.4508.01 and FWO. 
K. Hughes thanks the NSF for
support through the Graduate Research Fellowship Program Award DGE-1746045. A. Connolly thanks the NSF for
Award 1806923 and 2209588, and also acknowledges the Ohio Supercomputer Center. S. A. Wissel thanks the NSF for support through CAREER Award 2033500.
A. Vieregg thanks the Sloan Foundation and the Research Corporation for Science Advancement, the Research Computing Center and the Kavli Institute for Cosmological Physics at the University of Chicago for the resources they provided. R. Nichol thanks the Leverhulme
Trust for their support. K.D. de Vries is supported by
European Research Council under the European Unions
Horizon research and innovation program (grant agreement 763 No 805486). D. Besson, I. Kravchenko, and D. Seckel thank the NSF for support through the IceCube EPSCoR Initiative (Award ID 2019597). M.S. Muzio thanks the NSF for support through the MPS-Ascend Postdoctoral Fellowship under Award 2138121. A. Bishop thanks the Belgian American Education Foundation for their Graduate Fellowship support.

\end{document}